\documentclass[a4paper,paper]{JHEP3}

\usepackage{epsfig,multicol}

\title{$\alpha'$-exact entropies for BPS and non-BPS extremal dyonic black holes
       in heterotic string theory from ten-dimensional supersymmetry}

\author{Predrag Dominis Prester and Tomislav Terzi\'{c}\\
        Department of Physics, University of Rijeka\\
        Omladinska 14, HR-51000, Croatia\\
        E-mail: \email{pprester@phy.uniri.hr}, \email{tterzic@phy.uniri.hr}} 

\preprint{ 
          }

\abstract{We calculate near-horizon solutions for four-dimensional 4-charge and
five-dimensional 3-charge black holes in heterotic string theory from the part of the 
ten-dimensional tree-level effective action which is connected to gravitational Chern-Simons
term by supersymmetry. We obtain that the entropies of large black holes exactly match
the $\alpha'$-exact statistical entropies obtained from microstate counting ($D=4$) and 
AdS/CFT correspondence ($D=5$). Especially interesting is that we obtain agreement for both 
BPS and non-BPS black holes, contrary to the case of $R^2$-truncated (four-derivative) 
actions ($D$-dimensional $\mathcal{N}=2$ off-shell supersymmetric or Gauss-Bonnet) were 
used, which give the entropies agreeing (at best) just for BPS black holes. The key property of 
the solutions, which enabled us to tackle the action containing infinite number of terms, is 
vanishing of the Riemann tensor $\overline{R}_{MNPQ}$ obtained from torsional connection defined 
with $\overline{\Gamma} = \Gamma - \frac{1}{2} H$. Moreover, if every monomial of the remaining 
part of the effective action would contain at least two Riemanns $\overline{R}_{MNPQ}$, it would 
trivially follow that our solutions are exact solutions of the full heterotic effective action 
in $D=10$. The above conjecture, which appeared (in this or stronger form) from time to time in 
the literature, has controversial status, but is supported by the most recent calculations of 
Richards (arXiv:0807.3453 [hep-th]). Agreement of our results for the entropies with the 
microscopic ones supports the conjecture. As for small black holes, our solutions in $D=5$ 
still have singular horizons.}

\begin{document}

\section{Introduction: Motivation and results}
\label{sec:intro}

One of the most exciting results of string theory is that it has offered true microscopic 
statistical derivation of the Bekenstein-Hawking entropy formula for (near-)extremal
black holes\footnote{Subscripts on entropy denote: BH = Bekenstein-Hawking, 
bh = black hole (Wald formula), stat = statistical from direct microstate counting, 
CFT = statistical from the dual CFT.}
\begin{equation} \label{BHef}
S_{\rm BH} = \frac{A}{4 G_N} = \ln \mathcal{N}_{\rm micro} = S_{\rm stat}
\end{equation}
where $A$ is the area of the black hole, $G_N$ effective Newton constant, and 
$\mathcal{N}_{\rm micro}$ number of stringy microstates corresponding to the particular
black hole configuration.
Strictly speaking, (\ref{BHef}) is valid only asymptotically in the regime of large black holes.
The formula receives both the "classical" corrections due to the finite size of the string 
($\alpha'$-corrections) and "quantum" corrections due to finite value of string coupling 
(loop or $g_s$-corrections).

The understanding of corrections appears to be invaluable for several reasons. (i)
\emph{Non-trivial check of statistical description} -- in (\ref{BHef}) two sides of equality are
calculated in different non-overlapping regimes so the comparison is indirect. Statistical
calculation is done in the regime in which space-time is approximately flat, while on the
gravity side one deals with black hole space-time (though in Bekenstein-Hawking limit with 
small curvature outside the horizon). Identification of stringy microstates as black holes is 
obtained through a comparison of set of quantum numbers (charges), and by use of 
supersymmetry non-renormalization theorems for BPS states or attractor mechanism for non-BPS
ones. If the equality $S_{\rm bh} = S_{\rm stat}$ survives inclusion of corrections, we can be more 
confident in such indirect identifications. (ii) \emph{New insights and understanding of string 
theory.} -- investigations of the black holes in string theory led us to some important concepts 
(AdS/CFT correspondence, OSV conjecture, attractor mechanism, and structure of effective actions, 
to name the few), and corrections are also playing important role. (iii) \emph{Small black holes}
 -- for some special values of charges lowest order solutions have singular horizon with vanishing 
area. It was shown on some explicit examples that $\alpha'$-corrections generally stretch the 
horizon and regularize solutions. Still, such black holes are special in the sense that they are 
string-sized, and a lot of effort has been put recently to understand their properties (see
\cite{Castro:2008ne} for a recent review).

Analyses of $\alpha'$-corrections generally appear to be more straightforward, not only from 
calculation side but also conceptually. Here one calculates solutions and entropies from
well defined (at least in principle) tree-level action by using a generalization of 
Bekenstein-Hawking formula called Wald formula. On the other hand, as $\alpha'$- and 
$g_s$-corrections are interconnected by some duality relations, there is no sharp
division between them.

In this paper we concentrate on exact $\alpha'$-corrections for two special cases of black 
holes in the heterotic string theory: 4-charge in $D=4$ ($S^1 \times S^1 \times T^4$ 
compactification) and 3-charge in $D=5$ dimensions ($S^1 \times T^4$). These black holes are 
convenient as on the one hand they are simple enough to be treated in different ways, on the 
other hand they show rich behavior and even some surprises. Lowest-order solutions were 
calculated in \cite{Cvetic:1995uj,Cvetic:1995bj}. As $\alpha'$-corrections introduce higher 
derivatives, which drastically complicate calculations, we shall consider only near-horizon 
behavior which has enough information for calculation of black hole entropy.\footnote{Existence 
of regular uncorrected large black hole solutions for all (nonvanishing) values of charges, 
combined with apparent uniqueness, makes us confident that our near-horizon solutions are 
connected to physical asymptotically flat black holes (avoiding possible problems like those 
described in \cite{Chen:2008hk}).}

The case of 4-charge 4-dimensional extremal black holes is better understood (see 
\cite{Sen:2007qy} for detailed review). Statistical entropy formula has been calculated 
$\alpha'$-exactly \cite{Sen:2007qy}, which in BPS case ($n,w,N',W' > 0$) is given by
\begin{equation} \label{e4c4db}
S_{\rm stat}^{\rm (BPS)} = 2\pi \sqrt{nw(N'W'+4)} \,,
\end{equation}
and in non-BPS case (for which as a representative we take $n<0$ and $w,N',W'>0$) by
\begin{equation} \label{e4c4dn}
S_{\rm stat}^{\rm (non-BPS)} = 2\pi \sqrt{|n|w(N'W'+2)} \,.
\end{equation}
On the gravity side, as effective low energy action has an infinite expansion even on tree-level,
$\alpha'$-exact calculation appeared intractable. For this reason calculations were performed
by using certain $R^2$ truncated actions, namely (i) off-shell $\mathcal{N}=2$ supersymmetric
action \cite{Lopes Cardoso:1998wt,Behrndt:2005he,Sahoo:2006rp}, (ii) action with Gauss-Bonnet 
term \cite{Sen:2005iz}. Interestingly, both of these \emph{incomplete} actions
lead to the black hole entropies $S_{\rm bh}$ which \emph{exactly} agree with statistical entropy 
in the BPS case (\ref{e4c4db}), but completely fail (already at $\alpha'^1$-order) to reproduce 
(\ref{e4c4dn}) in the non-BPS case. Latter result shows that both truncated actions are 
incomplete already at $R^2$ ($\alpha'^1$-) order. On the other hand, calculations performed by 
using AdS$_3$/CFT$_2$ correspondence gave results agreeing \emph{both} with BPS and non-BPS 
entropies \cite{Kutasov:1998zh,Kraus:2005vz}. Especially interesting is the more general analysis 
\cite{Kraus:2005vz,David:2007ak} which shows that 3-dimensional $\mathcal{N}=4$ 
supersymmetry\footnote{The result has been recently extended to $(0,2)$ supersymmetry 
\cite{Kaura:2008us}.} 
implies that the only terms in the action which are important are Chern-Simons terms, which are 
1-loop saturated, suggesting that the only important $\alpha'$-corrections should be of $R^2$ 
(4-derivative) type. This gives some understanding for the success of $R^2$-truncated actions, but 
does not explain why such \emph{incomplete} actions are succeeding to describe entropy for BPS 
black holes (and at the same time fail for non-BPS).

For the 3-charge 5-dimensional extremal black holes the situation is even more interesting. Here,
direct microstate counting which would give statistical entropy has not been
yet performed, but from calculation \cite{Kutasov:1998zh} based on AdS$_3$/CFT$_2$ 
correspondence one can obtain statistical entropy, which in the BPS case ($n,w,N > 0$) is
\begin{equation} \label{e3c5db}
S_{\rm CFT}^{\rm (BPS)} = 2\pi \sqrt{nw(N+2)} \,,
\end{equation}
and in the non-BPS case (for which as a representative we take $n<0$ and $w,N>0$)
\begin{equation} \label{e3c5dn}
S_{\rm CFT}^{\rm (non-BPS)} = 2\pi \sqrt{|n|wN} \,.
\end{equation}
From the gravity viewpoint, calculations of black hole entropy were again performed by using
$R^2$ truncated actions. Five-dimensional $\mathcal{N}=2$ off-shell supersymmetric action with 
$R^2$ terms, obtained in \cite{Hanaki:2006pj} by supersymmetrizing gravitational Chern-Simons 
term, again gives black hole entropy \cite{Castro:2007hc,Alishahiha:2007nn,Cvitan:2007en} which 
exactly agrees with CFT statistical entropy in BPS case (\ref{e3c5db}), and fails already at first order 
in the non-BPS case \cite{Cvitan:2007en,Cvitan:2007pk}.\footnote{We assume that in formulae from 
\cite{Cvitan:2007en,Cvitan:2007pk} one makes proper definition for number of NS5 branes, i.e., 
$N=m+1$, as explained later in section \ref{sec:3chbh}.} As for the action with pure Gauss-Bonnet
correction, now it gives agreement in BPS case only at $\alpha'^1$-order, and again completely 
fails in the non-BPS case. General arguments based on 3-dimensional $\mathcal{N}=4$ SUSY and 
AdS/CFT are still valid, but explicit calculation of central charges and entropy is still missing (one 
would need corresponding $R^2$-supergravity action in $D=6$, which is not known). 
Again, a mystery is why $R^2$-truncated actions which are incomplete already at first order, are 
giving agreement (exactly or perturbatively) with statistical entropy  (only) for BPS black holes. 
That there is no generic problem with gravity description of non-BPS black holes is shown by explicit
perturbative calculations (up to $\alpha'^2$-order) of black hole entropies which are in agreement 
with statistical entropies (\ref{e4c4db})-(\ref{e3c5dn}) \cite{Sahoo:2006pm,Cvitan:2007hu}.

Motivated by such puzzles, we committed ourselves to calculate near-horizon solutions and entropies 
of above mentioned extremal black holes directly from ten-dimensional heterotic effective action 
(and without using AdS/CFT conjecture). First, we show that starting from the action which contains all 
terms connected to gravitational Chern-Simons term by on-shell $\mathcal{N}=1$ supersymmetry 
\cite{Bergshoeff:1989de}, one obtains black hole entropies  which \emph{exactly} agree with 
statistical entropies, both in BPS and non-BPS cases (\ref{e4c4db})-(\ref{e3c5dn}). The key property 
of near-horizon solutions, which enabled treatment of the action which has infinite 
$\alpha'$-expansion, is that they satisfy the relation
\begin{equation} \label{trie0}
\overline{R}_{MNPQ} = 0 \,,
\end{equation}
where $\overline{R}_{MNPQ}$ is the 10-dimensional Riemann tensor calculated from the torsional
connection defined in (\ref{modcon}).

It is known that heterotic effective action contains additional (unambiguous) terms starting from 
$\alpha'^3$-order, which include "infamous" $R^4$-type term multiplied by $\zeta(3)$ number.
Due to irrational nature of $\zeta(3)$, and perturbative character of large black holes, it
is obvious that such terms should give vanishing contribution to the entropy, but direct 
argument for such vanishing was missing. Our results show that not only this term but all
terms which are not connected with gravitational Chern-Simons term should give vanishing
total contribution to the entropy. Though the knowledge of this sector of the action is
largely incomplete and only few terms were explicitly computed, all but one\footnote{Selection
of the references with calculations in accord with conjecture is 
\cite{Gross:1986mw,Policastro:2006vt,Richards:2008sa}. The one which finds violation of
conjecture is \cite{Frolov:2001xr}, but the most recent calculations of the same terms in the action 
are showing the opposite \cite{Richards:2008sa}. It would be interesting to clear this controversy.} 
of these computations are in accord with the conjecture \cite{Kehagias:1997cq} that affine 
connection and 2-form field $B_{MN}$ enter this part of the effective Lagrangian solely through 
the torsional Riemann tensor $\overline{R}_{MNPQ}$. The trivial corollary of the conjecture (if 
correct), combined with the property (\ref{trie0}), is that the additional part of the action 
indeed \emph{does not make contribution to the near-horizon solution and entropy}, as expected. 
Turning the argument around, it could be said that our results support the conjecture. We 
emphasize that weaker assumptions would be enough for our purpose, e.g., that every monomial 
in the additional part of the action contains at least two powers of $\overline{R}_{MNPQ}$ (in fact, 
this was originally assumed in \cite{Metsaev:1987zx}). We mention that one trivial consequence of 
the conjecture is that for corresponding large black holes in type-II string theories 
near-horizon solutions and entropies are \emph{unaffected} by $\alpha'$-corrections, which is in 
accordance with OSV conjecture \cite{Ooguri:2004zv}.

Our results also shed some light on the puzzling aspects of small black holes, obtained when
magnetic charges ($N'$ and $W'$, or $N$) are taken to vanish. Such small black holes are
microscopically described by perturbative string (Dabholkar-Harvey) states for which statistical
entropy is asymptotically given in BPS case ($n,w>0$) by
\begin{equation} \label{esmallb}
S_{\rm stat}^{\rm (BPS)} = 4\pi \sqrt{nw} \,,
\end{equation}
and in non-BPS case (for which as a representative we take $n<0$, $w>0$) by
\begin{equation} \label{esmalln}
S_{\rm stat}^{\rm (non-BPS)} = 2\sqrt{2}\pi \sqrt{|n|w} \,.
\end{equation}
Now, if we naively put $N'=W'=N=0$ in formulae for large black holes, we see that in $D=4$ 
(\ref{e4c4db}) and (\ref{e4c4dn}) indeed give (\ref{esmallb}) and (\ref{esmalln}), while in $D=5$ 
(\ref{e3c5db}) and (\ref{e3c5dn}) give something different. Though we do not have the full
understanding of the small black hole limit, our results show that there is no real controversy
here. Though our solutions are singular in both dimensions, in $D=4$ singularity shows only in 
the sector which should decouple in the limit and apparently can be regularized by field 
redefinitions, while in $D=5$ solutions are completely singular and it is not obvious how to
regularize them.  

The outline of the paper goes as follows. In section \ref{sec:eahet} we recapitulate facts about
effective action of heterotic string theory in $D=10$, compactifications, and the way we handle
Chern-Simons term. In section \ref{sec:4chbh} we present our near-horizon solutions for 4-charge
extremal black holes in $D=4$, and in section \ref{sec:3chbh} the same for 3-charge black holes
in $D=5$ dimensions. In section \ref{sec:comm} we comment on connection with AdS/CFT 
constructions (sec. \ref{ssec:cft}), compare our solutions with those obtained from 
$R^2$-truncated actions (sec. \ref{ssec:prevan}), review the known facts and controversies on the 
structure of effective action (sec. \ref{ssec:torsion}), and comment the small black hole limit (sec. 
\ref{ssec:smallBH}).

\section{Effective action of heterotic string theories}
\label{sec:eahet}

\subsection{Ten-dimensional effective action}
\label{ssec:10Daction}

The 10-dimensional tree-level effective action of heterotic string theory has an infinite 
expansion in the string parameter $\alpha'$
\begin{equation} \label{lalphex}
\mathcal{S}^{(10)} = \int dx^{10} \sqrt{-G^{(10)}} \mathcal{L}^{(10)}
  = \sum_{n=0}^\infty \int dx^{10} \sqrt{-G^{(10)}} \mathcal{L}^{(10)}_n \;,
\end{equation}
where $G^{(10)}$ is a determinant of the 10-dimensional metric tensor $G^{(10)}_{MN}$.
As we are going to be interested in classical purely bosonic configurations which are uncharged 
under 1-form gauge fields, to simplify expressions we shall start from bosonic part of the 
tree-level effective action with 1-form gauge fields taken to be zero.
Then every $\mathcal{L}^{(10)}_n$ is a
function of the string metric $G^{(10)}_{MN}$, Riemann tensor $R^{(10)}_{MNPQ}$, dilaton 
$\Phi^{(10)}$, 3-form gauge field strength $H^{(10)}_{MNP}$ and the covariant derivatives of 
these fields. 10-dimensional space-time indices are denoted as $M,N,\ldots = 0,1,\ldots,9$. 
The term $\mathcal{L}^{(10)}_n$ has $2(n+1)$ derivatives, and is multiplied with
a factor of $\alpha'^n$.

Ten-dimensional Lagrangian can be decomposed in the following way
\begin{equation} \label{10dtl}
\mathcal{L}^{(10)} = \mathcal{L}^{(10)}_{01}
 + \Delta \mathcal{L}^{(10)}_{\rm CS} + \mathcal{L}^{(10)}_{\rm other} \;.
\end{equation}
The first term in (\ref{10dtl}), explicitly written, is
\begin{equation} \label{l100}
\mathcal{L}^{(10)}_{01} = \frac{e^{-2\Phi^{(10)}}}{16\pi G_{10}} \left[ R^{(10)} +
 4\left(\partial \Phi^{(10)}\right)^2 - \frac{1}{12} H^{(10)}_{MNP} H^{(10)MNP} \right] \;,
\end{equation}
where $G_{10}$ is 10-dimensional Newton constant.
3-form gauge field strength is not closed, but instead given by
\begin{equation} \label{hbcs}
H^{(10)}_{MNP} = \partial_M B^{(10)}_{NP} + \partial_N B^{(10)}_{PM}
 + \partial_P B^{(10)}_{MN} - 3 \alpha' \overline{\Omega}^{(10)}_{MNP} \;,
\end{equation}
where $\overline{\Omega}^{(10)}_{MNP}$ is the gravitational Chern-Simons form
\begin{equation} \label{gcs}
\overline{\Omega}^{(10)}_{MNP} = \frac{1}{2} \, \overline{\Gamma}^{(10)R}_{\quad\;\;\;\; MQ} \,
 \partial_N \overline{\Gamma}^{(10)Q}_{\quad\;\;\;\; PR} \, 
 + \, \frac{1}{3} \, \overline{\Gamma}^{(10)R}_{\quad\;\;\;\; MQ} \,
 \overline{\Gamma}^{(10)Q}_{\quad\;\;\;\; NS} \,
 \overline{\Gamma}^{(10)S}_{\quad\;\;\;\; PR} \;\; \mbox{(antisym. in $M,N,P$)}
\end{equation}
Bar on the geometric object means that it is calculated using a modified connection
\begin{equation} \label{modcon}
\overline{\Gamma}^{(10)P}_{\qquad MN} = \Gamma^{(10)P}_{\quad\;\; MN}
  - \frac{1}{2} H^{(10)P}_{\qquad MN}
\end{equation}
in which $H$ plays the role of a torsion. It is believed that Chern-Simons terms appear 
exclusively through Eq. (\ref{hbcs}).

If in (\ref{hbcs}) the Chern-Simons form $\overline{\Omega}^{(10)}_{MNP}$ would be absent,
then we would have $\mathcal{L}^{(10)}_{01} = \mathcal{L}^{(10)}_0$ in (\ref{lalphex}). Its 
presence introduces non-trivial $\alpha'$-corrections. Beside, as shown in \cite{Bergshoeff:1989de}, supersymmetrization (on-shell completion of $\mathcal{N}=1$ SUSY) of the Chern-Simons term 
introduces a (probably infinite) tower of terms in the effective action (with increasing number of 
derivatives), denoted by $\Delta \mathcal{L}^{(10)}_{\rm CS}$ in (\ref{10dtl}). The 
first two non-vanishing terms (in expansion in $\alpha'$) are\footnote{It was shown in 
\cite{Chemissany:2007he} that this effective action is to $\alpha'^1$-order 
equivalent (up to field redefinitions) to the one obtained in \cite{Metsaev:1987zx} directly from 
string amplitudes and sigma-model calculations.}
\begin{equation} \label{l101}
\Delta \mathcal{L}^{(10)}_{\mathrm{CS},1}  = \frac{\alpha'}{8}
 \frac{e^{-2\Phi^{(10)}}}{16\pi G_{10}} \overline{R}^{(10)}_{MNPQ} \overline{R}^{(10)MNPQ} 
\end{equation}
and
\begin{equation} \label{l103cs}
\Delta \mathcal{L}^{(10)}_{\mathrm{CS},3} = -\frac{\alpha'^3}{64}
 \frac{e^{-2\Phi^{(10)}}}{16\pi G_{10}}
 \left( 3\, T_{MNPQ}\, T^{MNPQ} + T_{MN}\, T^{MN} \right)
\end{equation}
where
\begin{equation} \label{tten}
T_{MNPQ} \equiv \overline{R}_{[MN}^{(10)\; RS} \, \overline{R}^{(10)}_{PQ]RS} \;,
 \qquad
T_{MN} \equiv \overline{R}_{MP}^{(10)\; QR}\, \overline{R}^{(10)P}_{\qquad NQR} \;.
\end{equation}

Though higher terms present in $\Delta \mathcal{L}^{(10)}_{\rm CS}$ were not 
explicitly constructed, it was argued in \cite{Bergshoeff:1989de} that $\alpha'^n$ contribution 
should be a linear combination of monomials containing $n$ Riemann tensors 
$\overline{R}_{MNPQ}$ calculated from the connection with torsion as given in (\ref{modcon}). 
This is the key information for us. All black hole near-horizon solutions that we construct and 
analyze in the paper have the property that $\overline{R}_{MNPQ}$ evaluated on them 
\emph{vanishes}, which means that all these terms, including (\ref{l101}) and (\ref{l103cs}),
will be irrelevant in our calculations.

It is well-known that, beside terms connected with Chern-Simons term by supersymmetry 
(analyzed above), additional terms appear in the effective action starting from $\alpha'^3$ 
(8-derivative) order. In (\ref{10dtl}) we have denoted them with $\mathcal{L}^{(10)}_{\rm other}$. 
One well-known example is $R^4$-type term multiplied by $\zeta(3)$, which appears in all string 
theories. Unfortunately, the knowledge of structure of $\mathcal{L}^{(10)}_{\rm other}$
is currently highly limited, and only few terms have been unambiguously calculated.

From now on, we are going to neglect contributions coming from $\mathcal{L}^{(10)}_{\rm other}$. 
One motivation is following from AdS$_3$/CFT$_2$ correspondence and anomaly inflow 
arguments of \cite{Kraus:2005vz}. There was argued (from 3-dimensional perspective) that for 
geometries having $AdS_3$ factor only Chern-Simons terms are important for calculations of 
central charges (from which one can calculate the black hole entropy). 
$\mathcal{L}^{(10)}_{\rm other}$ neither contains Chern-Simons terms nor is connected by 
supersymmetry to them, it should be irrelevant in such calculations. AdS$_3$/CFT$_2$ argument 
is sometimes used to explain successes of $R^2$-truncated actions (supersymmetric and/or 
Gauss-Bonnet) in calculations of entropies of BPS black holes in $D=4$ and 5. However, as the 
entropies obtained for non-BPS black holes fail to reproduce statistical entropies, we would like to 
avoid depending just on that argument (notice that static extremal black holes have 
$AdS_2 \times S^1$ (instead of $AdS_3$) factor, so the AdS$_3$/CFT$_2$ argument does not 
apply directly).

In fact, there is a more direct argument. If it happens that $\mathcal{L}^{(10)}_{\rm other}$ could 
be written in such a way that every monomial in it contains two powers of $\overline{R}_{MNPQ}$, 
then it would be irrelevant for our calculations and our results would be undoubtedly 
$\alpha'$-exact. The argument is the same as the one we used for 
$\Delta \mathcal{L}^{(10)}_{\rm CS}$ two paragraphs above. Indeed, this property was conjectured 
long time ago, see e.g., \cite{Metsaev:1987zx}. There is a stronger form of the conjecture, which 
claims that $\mathcal{L}^{(10)}_{\rm other}$ is purely composed of ($G^{MN}$ contracted) 
products of $\overline{R}_{MNPQ}$, see, e.g., \cite{Kehagias:1997cq}. Though the current status 
of the conjecture appears to be somewhat controversial -- it was disputed in 
\cite{Frolov:2001xr,Peeters:2001ub}, but the most recent detailed calculations 
\cite{Richards:2008sa} of some 8-derivative corrections (some of them recalculating the ones from 
\cite{Frolov:2001xr,Peeters:2001ub}) are giving results \emph{in agreement} with the strong form 
of the conjecture. We postpone discussion on this interesting topic until section \ref{ssec:torsion}. 
We can say that results of this paper are in agreement with the conjecture (at least in weak form).

\subsection{Manipulating Chern-Simons terms in $D=6$}
\label{ssec:cs6d}

All configurations that we analyze in this paper have four spatial dimensions compactified 
on torus $T^4$, and are uncharged under Kaluza-Klein 1-form gauge fields originating from four
compactified dimensions. Taking from the start that corresponding gauge fields 
vanish\footnote{Such truncation is expected to be consistent.} one obtains that the effective 
action is the same as in the section \ref{ssec:10Daction}, but now considering all fields and 
variables to be 6-dimensional. Effectively, one just has to replace everywhere $(10)$ with 
$(6)$ and take indices corresponding to 6-dimensional space-time, i.e., 
$M,N,\ldots = 0,1,\ldots,5$). To shorten the expressions, we immediately fix the values of
Newton constant and $\alpha'$, which in our normalization take values 
$G_6=2$ and $\alpha'=16$.

Appearance of gravitational Chern-Simons term in (\ref{hbcs}) introduces two problems. One is 
that it introduces in the action terms which are not manifestly diff-covariant, and that prevents 
direct use of Sen's entropy function formalism. A second problem is that due to (\ref{modcon}) 
and (\ref{hbcs}) Chern-Simons term is mixed in a complicated way with other 
$\alpha'$-corrections. We handle these problems by using the following two-step procedure 
(introduced in \cite{Sahoo:2006pm}).

First, we introduce an additional 3-form $K^{(6)}=dC^{(6)}$ and put a theory in a classically
equivalent form in which Lagrangian is given by
\begin{eqnarray}
\sqrt{-G^{(6)}} \widetilde{\mathcal{L}}^{(6)} &=&
 \sqrt{-G^{(6)}} \mathcal{L}^{(6)}
 + \frac{1}{(24\pi)^2} \epsilon^{MNPQRS} K^{(6)}_{MNP} H^{(6)}_{QRS}
\nonumber \\
&& + \frac{3\alpha'}{(24\pi)^2} \epsilon^{MNPQRS} 
 K^{(6)}_{MNP} \overline{\Omega}^{(6)}_{QRS} \;,
\label{gnewact}
\end{eqnarray}
and where now $H^{(6)}_{MNP}$ should not be treated as a gauge field strength
but as an auxiliary 3-form. Antisymmetric tensor density $\epsilon^{MNPQRS}$ is defined 
by $\epsilon^{012345}=1$. As a result, Chern-Simons term is now isolated
as a single $\alpha'^1$-correction, in a way which will eventually allow us to write it in a 
manifestly covariant form. 

Before passing to a second step of the procedure from \cite{Sahoo:2006pm}, we need to 
isolate in (\ref{gnewact}) ordinary Chern-Simons term $\Omega^{(6)}$ (obtained from
standard Levi-Civita connection) from the rest by using (\ref{modcon}). The result is
\cite{Chemissany:2007he}
\begin{equation} \label{bcscs}
\overline{\Omega}^{(6)}_{MNP} = \Omega^{(6)}_{MNP} + \mathcal{A}^{(6)}_{MNP}
\end{equation}
where
\begin{eqnarray} \label{a6mnp}
 \mathcal{A}^{(6)}_{MNP} &=&
 \frac{1}{4} \partial_M \left( \Gamma^{(6)R}_{NQ} H^{(6)Q}_{RP} \right)
 + \frac{1}{8} H^{(6)R}_{MQ} \nabla_N H^{(6)Q}_{RP}
 - \frac{1}{4} R^{(6)\; QR}_{MN} H^{(6)}_{PQR} \nonumber \\
&& + \frac{1}{24} H^{(6)R}_{MQ} H^{(6)S}_{NR} H^{(6)Q}_{PS} \quad
 \mbox{(antisymmetrized in $M,N,P$).}
\end{eqnarray}
Notice that when (\ref{a6mnp}) is plugged in (\ref{bcscs}), and this into (\ref{gnewact}),
which is then integrated to obtain the action, contribution from the first term in 
(\ref{a6mnp}) will, after partial integration, have a factor $dK^{(6)}$ which vanishes 
because $K^{(6)}$ is by definition exact form. We now see that $\mathcal{A}^{(6)}$ gives 
manifestly covariant contribution to the action.

Now we are ready to write 6-dimensional action
\begin{equation} \label{6dea}
\mathcal{S}^{(6)} = \int dx^{6} \sqrt{-G^{(6)}} \widetilde{\mathcal{L}}^{(6)} 
\end{equation}
in the form we are going to use extensively in the paper.
Using (\ref{10dtl}) and the above analysis, Lagrangian can be written in the following form
\begin{equation} \label{6dtl}
\widetilde{\mathcal{L}}^{(6)} = \widetilde{\mathcal{L}}^{(6)}_0
 + \widetilde{\mathcal{L}}^{(6)\prime}_1 + \widetilde{\mathcal{L}}^{(6)\prime\prime}_1
 + \Delta \mathcal{L}^{(6)}_{\rm CS} + \mathcal{L}^{(6)}_{\rm other} \;.
\end{equation}
First term is lowest order ($\alpha'^0$) contribution given by (\ref{l100})
\begin{equation} \label{6dtl0}
\widetilde{\mathcal{L}}^{(6)}_0 = \frac{e^{-2\Phi^{(6)}}}{32\pi} \left[ R^{(6)} +
 4\left(\partial \Phi^{(6)} \right)^2 - \frac{1}{12} H^{(6)}_{MNP} H^{(6)MNP} \right]
 + \frac{\epsilon^{MNPQRS}}{(24\pi)^2 \sqrt{-G^{(6)}}} K^{(6)}_{MNP} H^{(6)}_{QRS}
\end{equation}
For later convenience we have separated first-order terms in three parts. One is given by
\begin{equation} \label{6dtl1p}
\widetilde{\mathcal{L}}^{(6)\prime}_1 = \frac{\epsilon^{MNPQRS}}{12\pi^2 \sqrt{-G^{(6)}}}
 K^{(6)}_{MNP} \left( \frac{1}{8} H^{(6)U}_{QT} \nabla_R H^{(6)T}_{US}
 - \frac{1}{4} R^{(6)\; TU}_{QR} H^{(6)}_{STU}
 + \frac{1}{24} H^{(6)U}_{QT} H^{(6)V}_{RU} H^{(6)T}_{SV} \right)
\end{equation}
The second part, which contains gravitational Chern-Simons term and is not manifestly 
covariant, is given by
\begin{equation} \label{6dtl1pp}
\widetilde{\mathcal{L}}^{(6)\prime\prime}_1 = \frac{\epsilon^{MNPQRS}}{12\pi^2 \sqrt{-G^{(6)}}}
 K^{(6)}_{MNP} \Omega^{(6)}_{QRS} \;.
\end{equation}
Finally, the third part is contained in $\Delta \mathcal{L}^{(6)}_{\rm CS}$ (\ref{l101}). In 
\cite{Sahoo:2006pm} it was shown how to rewrite (\ref{6dtl1pp}) in the manifestly covariant form 
for the particular type of the backgrounds which includes those we shall analyze in this paper.

Following the discussion from the previous section, we shall first calculate near-horizon 
solutions by ignoring $\Delta \mathcal{L}^{(6)}_{\rm CS}$ and $\mathcal{L}^{(6)}_{\rm other}$ 
in the effective action (\ref{6dtl}). As all solutions that we obtain satisfy 
$\overline{R}_{MNPQ} = 0$, from the conjecture that these terms can be written in such a way 
that every monomial in them contains two powers of $\overline{R}_{MNPQ}$ directly follows that 
they are irrelevant for our calculations. As already discussed, for 
$\Delta \mathcal{L}^{(6)}_{\rm CS}$ validity of the conjecture was argued in 
\cite{Bergshoeff:1989de}, while for $\mathcal{L}^{(6)}_{\rm other}$ the situation is still unclear, 
though explicit calculations are apparently supporting it (for more details see section 
\ref{ssec:torsion}). 

All in all, we shall start from the reduced action with Lagrangian given by
\begin{equation} \label{6dtlred}
\widetilde{\mathcal{L}}^{(6)}_{\rm red} = \widetilde{\mathcal{L}}^{(6)}_0
 + \widetilde{\mathcal{L}}^{(6)\prime}_1 + \widetilde{\mathcal{L}}^{(6)\prime\prime}_1 \;,
\end{equation}
and check if the near horizon solutions satisfy the condition $\overline{R}_{MNPQ} = 0$. If
this is satisfied, it follows immediately that they are also solutions of the action with Lagrangian
\begin{equation} \label{6dtlsusy}
\widetilde{\mathcal{L}}^{(6)}_{\rm susy} = \widetilde{\mathcal{L}}^{(6)}_{\rm red}
 + \Delta\mathcal{L}^{(6)}_{\rm CS} \;,
\end{equation}
and, under the above mentioned assumption on $\mathcal{L}^{(6)}_{\rm other}$, of the
full heterotic action (\ref{6dtl}).

To avoid confusion, we emphasize that action (\ref{6dtlred}) is equivalent (for backgrounds 
satisfying $\overline{R}_{MNPQ} = 0$)  to the 4-derivative action used in \cite{Sahoo:2006pm} up 
to $\alpha'^1$-order, but that field redefinition which connects them introduces non-vanishing 
higher $\alpha'$-corrections. This was confirmed in \cite{Cvitan:2007hu}, where it was shown that 
action from \cite{Sahoo:2006pm} must be supplemented with higher-derivative terms, as indeed it 
is expected on general grounds.

\subsection{Compactification to $D<6$}
\label{ssec:Dcomp}

Our main interest are black holes in $D=5$ and $D=4$ dimensions, so we 
consider further compactification on $(6-D)$ circles $S^1$. Using the standard
Kaluza-Klein compactification we obtain $D$-dimensional fields
$G_{\mu\nu}$, $C_{\mu\nu}$, $\Phi$, $\widehat{G}_{mn}$,
$\widehat{C}_{mn}$ and $A_\mu^{(i)}$ ($0\le\mu,\nu\le D-1$,
$D\le m,n\le 5$, $1\le i\le 2(6-D)$): 
\begin{eqnarray} \label{d6dD}
&& \widehat{G}_{mn} = G^{(6)}_{mn}\,, \quad
 \widehat{G}^{mn} = (\widehat{G}^{-1})^{mn} \,, \quad
 \widehat{C}_{mn} = C^{(6)}_{mn}\,,
\nonumber \\
&& A^{(m-D+1)}_\mu = \frac{1}{2} \widehat{G}^{nm} G^{(6)}_{n\mu} \,,
 \quad
 A^{(m-2D+7)}_\mu = \frac{1}{2} C^{(6)}_{m\mu}
 - \widehat{C}_{mn} A^{(n-D+1)}_\mu \, ,
\nonumber \\
&& G_{\mu\nu} = G^{(6)}_{\mu\nu} - \widehat{G}^{mn} G^{(6)}_{m\mu}
 G^{(6)}_{n\nu}\, ,
\nonumber \\
&& C_{\mu\nu} = C^{(6)}_{\mu\nu} - 4 \widehat{C}_{mn}
 A^{(m-D+1)}_\mu A^{(n-D+1)}_\nu
 - 2(A^{(m-D+1)}_\mu A^{(m-2D+7)}_\nu
 - A^{(m-D+1)}_\nu A^{(m-2D+7)}_\mu)
\nonumber \\
&& \Phi = \Phi^{(6)} - \frac{1}{2} \ln \mathcal{V}_{6-D} \,,
\end{eqnarray}
There is also (now auxiliary) field $H^{(6)}_{MNP}$ which produces
$D$-dimensional fields $H_{\mu\nu\rho}$, $H_{\mu\nu m}$, 
$H_{\mu mn}$ and $H_{mnp}$. 
As in \cite{Sahoo:2006pm}, we take for the circle coordinates 
$0\le x^m<2\pi\sqrt{\alpha'}=8\pi$, so that the volume 
$\mathcal{V}_{6-D}$ is
\begin{equation} \label{volum}
\mathcal{V}_{6-D} = (8\pi)^{6-D} \sqrt{\widehat{G}} \;.
\end{equation}
The gauge invariant field strengths associated with
$A_\mu^{(i)}$ and $C_{\mu\nu}$ are
\begin{equation} \label{eag2a}
F^{(i)}_{\mu\nu} =
 \partial_\mu A^{(i)}_\nu - \partial_\nu A^{(i)}_\mu \, ,
 \qquad 1\le i,j\le 2(6-D) \, ,
\end{equation}
\begin{equation} \label{eag2b}
K_{\mu\nu\rho} = \left( \partial_\mu C_{\nu\rho}
 + 2 A_\mu^{(i)} L_{ij} F^{(j)}_{\nu\rho} \right)
 + \hbox{cyclic permutations of $\mu$, $\nu$, $\rho$} \, ,
\end{equation}
where
\begin{equation} \label{edefl}
L = \pmatrix{ 0 & I_{6-D} \cr I_{6-D} & 0} \, ,
\end{equation}
$I_{6-D}$ being a $(6-D)$-dimensional identity matrix.

For the black holes we are interested in, we have
\begin{equation}
A_\mu^{(i)} L_{ij} F_{\nu\rho}^{(j)} = 0 \;.
\end{equation}

Normally, the next step would be to perform the Kaluza-Klein reduction on the 6-di\-men\-sional 
low-energy effective action to obtain a $D$-dimensional effective action, which can be quite 
complicated. In \cite{Sahoo:2006pm} a simpler procedure is suggested -- one goes to $D$ 
dimensions just to use the symmetries of the action to construct an ansatz for the background 
($AdS_2\times S^{D-2}$ in our case) and then performs an uplift to 6 dimensions (by inverting 
(\ref{d6dD})) where the action is simpler and calculations are easier. We shall follow this logic here.

\section{4-dimensional 4-charge black holes in heterotic theory}
\label{sec:4chbh}

Here we consider the 4-dimensional 4-charge extremal black holes 
appearing in the heterotic string theory compactified on 
$T^4\times S^1\times S^1$. 
One can obtain an effective 4-dimensional theory by putting $D=4$ in
(\ref{d6dD}) (using the formulation of the 6-dimensional action from section 
\ref{ssec:cs6d}) and taking as non-vanishing only the following fields: string 
metric $G_{\mu\nu}$, dilaton $\Phi$, moduli $T_1=(\widehat{G}_{44})^{1/2}$ and 
$T_2=(\widehat{G}_{55})^{1/2}$, four Kaluza-Klein gauge fields 
$A_\mu^{(i)}$ ($0\le\mu,\nu\le3$, $1\le i\le 4$) coming from 
$G^{(6)}_{MN}$ and 2-form potential $C^{(6)}_{MN}$, and two auxiliary 2-forms 
$D_{\mu\nu}^{(n)}$ ($n=1,2$) coming from $H^{(6)}_{MNP}$ (which is now, as 
explained in section \ref{ssec:cs6d}, an auxiliary field). 

The black holes we are interested in are charged purely electrically with respect 
to $A_\mu^{(1)}$ and $A_\mu^{(3)}$, and purely magnetically with respect to 
$A_\mu^{(2)}$ and $A_\mu^{(4)}$. From heterotic string theory viewpoint, these 
black holes should correspond to 4-charge states in which, beside fundamental 
string wound around one $S^1$ circle (with coordinate $x^4$), and with nonvanishing 
momentum on it, there are also Kaluza-Klein and H-monopoles (NS5-branes) wound 
around the same $S^1$ and $T^4$ (with "nut" on second $S^1$).

For extremal black holes one expects $AdS_2\times S^2$ near-horizon geometry 
\cite{Kunduri:2007vf,Astefanesei:2007bf,Goldstein:2007km} which in the present case is given 
by:
\begin{eqnarray} \label{d6d4}
&& ds^2 \equiv G_{\mu\nu} dx^\mu dx^\nu = v_1 \left( -r^2 dt^2 + {dr^2\over r^2} \right)
 + v_2 (d\theta^2 + \sin^2\theta d\phi^2)\, , \nonumber \\
&& e^{-2\Phi} = u_S \,,\qquad T_1 = u_1 \,,\qquad T_2 = u_2 \,, \nonumber \\
&& F^{(1)}_{rt} = \widetilde{e}_1, \qquad F^{(3)}_{rt}
 = \frac{\widetilde{e}_3}{16} \,, \qquad
 F^{(2)}_{\theta\phi} = \frac{\widetilde{p}_2}{4\pi} \sin\theta \,,
 \qquad F^{(4)}_{\theta\phi} = \frac{\widetilde{p}_4}{64\pi}\sin\theta \,,\nonumber \\
&& D^{(1)\,rt} = \frac{2\,u_1^2\, h_1}{v_1 v_2 u_S} \,, \qquad
 D^{(2)\,\theta\phi} = -\frac{8\pi\, u_2^2\,h_2}{v_1 v_2 u_S \sin\theta} \,.
\end{eqnarray}
Here $v_1$, $v_2$, $u_S$, $u_n$, $\widetilde{e}_i$ and $h_n$ ($n=1,2$, $i=1,\ldots,4$) 
are unknown variables fixed by equations of motion and values of electric charges 
$\widetilde{q}_{1,3}$. Somewhat unusual normalization for $h_n$ is introduced for later 
convenience.

For the backgrounds obeying the full group of symmetries of $AdS_2 \times S^{D-2}$ space,
the most efficient way for finding solutions is to use Sen's entropy function formalism
developed in \cite{Sen:2005wa}. One defines the entropy function as
\begin{equation} \label{entf}
\mathcal{E} = 2\pi \left( \sum_{I} \widetilde{q}_I \, \widetilde{e}_I
 - \int_{S^2} \sqrt{-G} \, \widetilde{\mathcal{L}} \right) \;,
\end{equation}
where $\widetilde{q}_I$ are electric charges. On the right hand side one puts the 
near-horizon background (\ref{d6d4}) and integrates over the surface of the horizon 
(which is a 2-dimensional sphere in the present case). $\widetilde{\mathcal{L}}$ is the 
effective Lagrangian in four dimensions. Equations of motion are obtained by extremizing
the entropy function (\ref{entf}) over variables $\{\varphi_a\}=\{v_1,v_2, u_S, u_n, 
\widetilde{e}_i,h_n\}$,
\begin{equation} \label{geneom}
0 = \frac{\partial\mathcal{E}}{\partial\varphi_a} \,
 \Bigg|_{\varphi=\bar{\varphi}} \;.
\end{equation}
One obtains a system of algebraic equations. Finally, the black hole entropy, defined by Wald
formula \cite{Wald}, is given by the value of the entropy function at the extremum
\begin{equation} \label{bhgen}
S_{\rm bh} = \mathcal{E}(\bar{\varphi}) \;,
\end{equation}
which is a function of electric and magnetic charges only.

Instead of calculating $\widetilde{\mathcal{L}}$ by doing dimensional reduction from six 
to four dimensions, it is much easier to perform calculation of entropy function $\mathcal{E}$
directly in six dimensions were we already know the action. For this, we have to lift the
background to six dimensions, which for (\ref{d6d4}) gives
\begin{eqnarray} \label{d4d6}
&& ds_6^2 \equiv G^{(6)}_{MN} dx^M dx^N
 = ds^2 + u_1^2 \left( dx^4 + 2\widetilde{e}_1 rdt \right)^2
 + u_2^2 \left( dx^5 - \frac{\widetilde{p}_2}{2\pi} \cos\theta\, d\phi \right)^2 \,,
\nonumber \\
&& K^{(6)}_{tr4} = \frac{\widetilde{e}_3}{8} \,, \qquad
 K^{(6)}_{\theta\phi 5} = -\frac{\widetilde{p}_4}{32\pi} \sin\theta \,,
\nonumber \\
&& H^{(6)tr4} = \frac{4\,h_1}{v_1 v_2 u_S} \,, \qquad
 H^{(6)\theta\phi 5} = \frac{16\pi\,h_2}{v_1 v_2 u_S \sin\theta} \,,
\nonumber \\
&& e^{-2\Phi^{(6)}} =  \frac{u_S}{64\pi^2\, u_1 u_2} \,.
\end{eqnarray}
Instead of $\widetilde{\mathcal{L}}$ and $G$ we now use in (\ref{entf}) the six dimensional 
Lagrangian $\widetilde{\mathcal{L}}^{(6)}$ given in (\ref{6dtl})-(\ref{6dtl1pp}) and the 
determinant $G^{(6)}$
\begin{equation} \label{entf6d}
\mathcal{E} = 2\pi \left( \sum_{I} \widetilde{q}_I \, \widetilde{e}_I
 - \int_{S^2} \sqrt{-G^{(6)}} \, \widetilde{\mathcal{L}}^{(6)} \right) \;.
\end{equation}
This is obviously equivalent to (\ref{entf}).

As we discussed in sections \ref{ssec:10Daction} and \ref{ssec:cs6d}, we concentrate on the part 
of the action connected by 10-dimensional supersymmetry with Chern-Simons term (obtained by 
neglecting $\mathcal{L}^{(6)}_{\rm other}$ in (\ref{6dtl})). For the moment we also neglect 
$\Delta\mathcal{L}^{(6)}_{\rm CS}$, for which we show a posteriori that it does not contribute to 
the near-horizon solutions and the entropies. This means that we start with the reduced Lagrangian 
$\widetilde{\mathcal{L}}^{(6)}_{\rm red}$ defined by (\ref{6dtlred}), (\ref{6dtl0}), (\ref{6dtl1p}) and 
(\ref{6dtl1pp}). Putting (\ref{d4d6}) in (\ref{6dtlred}), and then this into the entropy function 
(\ref{entf6d}), we obtain
\begin{equation} \label{entfcs}
\mathcal{E} = \mathcal{E}_0 + \mathcal{E}'_1 + \mathcal{E}''_1 \,,
\end{equation}
where
\begin{eqnarray} \label{ef4c0}
\mathcal{E}_0 &=& 2\pi \left[ \widetilde{q}_1 \widetilde{e}_1
 + \widetilde{q}_3 \widetilde{e}_3
 - \int d\theta\, d\phi\, dx^4 dx^5 \sqrt{-G^{(6)}} \widetilde{\mathcal{L}}^{(6)}_0 \right]
\nonumber \\
 &=& 2\pi \left[ \widetilde{q}_1 \widetilde{e}_1 + \widetilde{q}_3 \widetilde{e}_3
  - \frac{1}{8} v_1 v_2 u_S \left( -\frac{2}{v_1} + \frac{2}{v_2}
 + \frac{2 u_1^2\, \widetilde{e}_1^2}{v_1^2}
 + \frac{128\pi^2 u_2^2 h_2 (2\widetilde{e}_3 - h_2)}{v_1^2\, u_S^2}
 \right. \right.
\nonumber \\
&& \left. \left. \qquad
 - \frac{u_2^2\, \widetilde{p}_2^2}{8\pi^2 v_2^2}
 - \frac{8 u_1^2 h_1 (2\widetilde{p}_4 - h_1)}{v_2^2\, u_S^2} \right)
 \right] \,,
\end{eqnarray}
and
\begin{eqnarray} \label{ef4c1}
\mathcal{E}'_1 &=& -2\pi \int d\theta\, d\phi\, dx^4 dx^5 \sqrt{-G^{(6)}}
 \widetilde{\mathcal{L}}^{(6)\prime}_1
\nonumber \\
&=& -4\pi v_1 v_2 u_S \left( \frac{8192\pi^4 u_2^4 \widetilde{e}_3 h_2^3}{v_1^4\, u_S^4}
 + \frac{8 u_2^4 \widetilde{e}_3 h_2 \widetilde{p}_2^2}{v_1^2\, v_2^2\, u_S^2} 
 - \frac{128\pi^2 u_2^2 \widetilde{e}_3 h_2}{v_1^2\, v_2\, u_S^2} \right.
\nonumber \\
&& \qquad\qquad\qquad \left.  + \frac{32 u_1^4 \widetilde{p}_4 h_1^3}{v_2^4\, u_S^4}
 + \frac{8 u_1^4 \widetilde{e}_1^2 h_1 \widetilde{p}_4}{v_1^2\, v_2^2\, u_S^2}
 - \frac{8 u_1^2 \widetilde{p}_4 h_1}{v_1\, v_2^2\, u_S^2}
 \right) \,.
\end{eqnarray}
With $\mathcal{E}''_1$ the situation is a bit tricky because of the presence of Chern-Simons
density in (\ref{6dtl1pp}). This means that $\widetilde{\mathcal{L}}^{(6)\prime\prime}_1$ is 
not manifestly diffeomorphism covariant, and one cannot apply directly Sen's entropy function 
formalism. Fortunately, this problem was solved in \cite{Sahoo:2006pm} where it was shown how 
for the class of the metrics, to which (\ref{d4d6}) belongs, one can write $\mathcal{E}''_1$ 
in a manifestly covariant form. We simply copy the final result (eq. (3.33) of 
\cite{Sahoo:2006pm})
\begin{eqnarray} \label{ef4c12}
\mathcal{E}''_1 &=& -2\pi \int d\theta\, d\phi\, dx^4 dx^5 \sqrt{-G^{(6)}}
 \widetilde{\mathcal{L}}^{(6)\prime\prime}_1
\nonumber \\ 
&=& - (8\pi)^2 \left[ \frac{\widetilde{p}_4}{4\pi}
 \left( \frac{u_1^2}{v_1} \widetilde{e}_1 - 2 \frac{u_1^4}{v_1^2}\,\widetilde{e}_1^3 \right)
 + \widetilde{e}_3 \, \left( \frac{u_2^2}{v_2} \, \frac{\widetilde{p}_2}{4\pi}
 - 2 \frac{u_2^4}{v_2^2} \, \left( \frac{\widetilde{p}_2}{4\pi} \right)^3 \right) \right] \,.
\end{eqnarray}

We are now ready to find near-horizon solutions, by solving the system (\ref{geneom}), and 
black hole entropy from (\ref{bhgen}). As we want to compare the results with the statistical
entropy obtained in string theory by counting of microstates, it is convenient to express 
charges $(\widetilde{q},\widetilde{p})$ in terms of (integer valued) charges naturally 
appearing in the string theory. In \cite{Sahoo:2006pm} it was shown that in the present case 
the correspondence is given by
\begin{equation} \label{chnorm}
\widetilde{q}_1 = \frac{n}{2} \,,\qquad \widetilde{p}_2 = 4\pi N' \,,\qquad
 \widetilde{q}_3 = - 4\pi W' \,,\qquad \widetilde{p}_4 = - \frac{w}{2} \,,
\end{equation}
where $n$ and $w$ are momentum and winding number of string wound along circle $x^4$, 
and $N'$ and $W'$ are Kaluza-Klein monopole and H-monopole charges associated with 
the circle $x^5$.

Using (\ref{entfcs})-(\ref{chnorm}) in (\ref{geneom}), we obtain quite a complicated algebraic
system, naively not expected to be solvable analytically. Amazingly, we have found analytic 
near-horizon solutions for all values of charges, corresponding to BPS and non-BPS black 
holes.\footnote{The way we constructed solutions was indirect - we managed to conjecture them 
from perturbative calculations (which we did up to $\alpha'^4$), and then checked them by putting 
into exact equations. For some special sets of charges we then numerically checked that there are 
no other physically acceptable solutions.} While in BPS case analytic solutions are expected 
because one can use BPS conditions to drastically simplify calculations, in non-BPS case in 
theories which involve higher-derivative corrections analytic solutions are typically not 
known\footnote{However,  one exception can be found in \cite{Cvitan:2007en}.}. Indeed, this is 
exactly what was observed when 4-dimensional $R^2$ supersymmetric action was used, in which
case only perturbative analysis was possible \cite{Sen:2007qy}.

For clarity of presentation, we take $w,N',W'>0$. Then $n>0$ ($n<0$) corresponds to 
BPS (non-BPS) black holes, respectively. In the BPS case near-horizon solutions are given 
by
\begin{eqnarray} \label{4solb}
&& v_1 = v_2 = 4(N'W'+2) \,,\qquad u_S = \sqrt{\frac{nw}{N'W'+4}} \,,
\nonumber \\
&& u_1 = \sqrt{\frac{n(N'W'+2)}{w(N'W'+4)}} \,,\qquad
 u_2 = \sqrt{\frac{W'}{N'} \left(1+\frac{2}{N'W'}\right)} \,,
\\
&& \widetilde{e}_1 = \frac{1}{n} \sqrt{nw(N'W'+4)} \,,\qquad
 \widetilde{e}_3 = h_2 = - \frac{N'}{8\pi} \sqrt{\frac{nw}{N'W'+4}} \,,\qquad
 h_1= - \frac{w}{2} \,.
\nonumber
\end{eqnarray}
For the entropy we obtain
\begin{equation} \label{4entb}
S_{\rm bh}^{\rm BPS} = 2\pi \sqrt{nw(N'W'+4)} \,.
\end{equation}
This is exactly what one obtains by microstate counting in string theory (\ref{e4c4db}), in
the limit $nw \gg N'W'$, which corresponds to tree-level approximation on gravity side.

In the non-BPS case we obtain
\begin{eqnarray} \label{4soln}
&& v_1 = v_2 = 4(N'W'+2) \,,\qquad u_S = \sqrt{\frac{|n|w}{N'W'+2}} \,,
\nonumber \\
&& u_1 = \sqrt{\frac{|n|}{w}} \,,\qquad\qquad\qquad\quad
 u_2 = \sqrt{\frac{W'}{N'} \left(1+\frac{2}{N'W'}\right)} \,,
\\
&& \widetilde{e}_1 = \frac{1}{n} \sqrt{|n|w(N'W'+2)} \,,\qquad
 \widetilde{e}_3 = h_2 = - \frac{N'}{8\pi} \sqrt{\frac{|n|w}{N'W'+2}} \,,\qquad
 h_1= - \frac{w}{2} \,.
\nonumber
\end{eqnarray}
For the entropy we obtain
\begin{equation} \label{4entn}
S_{\rm bh}^{\rm non-BPS} = 2\pi \sqrt{|n|w(N'W'+2)} \,.
\end{equation}
Again, agreement with statistical calculation in string theory (\ref{e4c4dn}) is exact in $\alpha'$.

Now we have to check that $\overline{R}^{(6)}_{MNPQ}$ vanishes when evaluated on our 
solutions. From (\ref{modcon}) one gets
\begin{equation} \label{modrie}
\overline{R}^{(6)M}_{\qquad NPQ} = R^{(6)M}_{\qquad NPQ} + \nabla_{[P} H^{(6)M}_{\qquad Q]N}
 - \frac{1}{2} H^{(6)M}_{\qquad R[P} H^{(6)R}_{\qquad Q]N} \;.
\end{equation}
It is easy to show that, both for BPS (\ref{4solb}) and non-BPS (\ref{4soln}) solutions,
6-dimensional background (\ref{d4d6}) has
\begin{equation} \label{modrie0}
\overline{R}^{(6)}_{MNPQ} = 0 \;.
\end{equation}
As explained in sections \ref{ssec:10Daction} and \ref{ssec:cs6d}, from this follows that
inclusion of the term $\Delta\mathcal{L}^{(6)}_{\rm CS}$ does not change neither the 
near-horizon solutions (\ref{4solb}) and (\ref{4soln}) nor the corresponding black hole 
entropies (\ref{4entb}) and (\ref{4entn}), which means that all our results would be
obtained if we started with the more complicated supersymmetric Lagrangian (\ref{6dtlsusy}), 
constructed by supersymmetrizing gravitational Chern-Simons term.

\section{5-dimensional 3-charge black holes in heterotic theory}
\label{sec:3chbh}

Here we consider the 5-dimensional spherically symmetric 3-charge extremal 
black holes which appear in the heterotic string theory compactified on
$T^4\times S^1$. One can obtain an effective 5-dimensional theory by putting $D=5$ in
(\ref{d6dD}) (again using the formulation of the 6-dimensional action from section 
\ref{ssec:cs6d}) and taking as non-vanishing only the following fields: string metric
$G_{\mu\nu}$, dilaton $\Phi$, modulus $T=(\widehat{G}_{55})^{1/2}$, two Kaluza-Klein
gauge fields $A_\mu^{(i)}$ ($0\le\mu,\nu\le4$, $1\le i\le 2$) coming from 
$G^{(6)}_{MN}$ and 2-form potential $C^{(6)}_{MN}$, the 2-form potential $C_{\mu\nu}$
with the strength $K_{\mu\nu\rho}$, one Kaluza-Klein auxiliary two form $D_{\mu\nu}$
coming from $H^{(6)}_{MNP}$, and auxiliary 3-form $H_{\mu\nu\rho}$. 

The black holes we are interested in are charged purely electrically with respect to 
$A_\mu^{(i)}$, and purely magnetically with respect to $K_{\mu\nu\rho}$. From the 
heterotic string theory viewpoint, these black holes should correspond to 3-charge
states in which, beside fundamental string wound around $S^1$ circle with nonvanishing
momentum on it, there are NS5-branes wrapped around $T^4\times S^1$.

For extremal black holes we now expect\footnote{In $D=5$ there is no explicit proof that 
extremal asymptotically flat black holes must have $AdS_2\times S^3$ near-horizon geometry.
However, for the large black holes analyzed here one knows that lowest order solutions, which
were fully constructed, have such near-horizon behavior, and from continuity one expects the
same when $\alpha'$-corrections are included. We note that the situation is not that clear for 
small black holes, which we shall discuss later.} $AdS_2\times S^3$ near-horizon geometry 
which in the present case is given by:
\begin{eqnarray} \label{d6d5}
&& ds^2 \equiv G_{\mu\nu} dx^\mu dx^\nu
 = v_1\left(-r^2 dt^2 + {dr^2\over r^2}\right) + v_2 d\Omega_3 \,, 
\nonumber \\
&& F^{(1)}_{rt} = \widetilde{e}_1, \qquad
 F^{(2)}_{rt} = \frac{\widetilde{e_2}}{4} \,, \qquad
K_{234} = \frac{\widetilde{p}}{4} \sqrt{g_3} \,,
\nonumber \\
&& D^{rt} = \frac{2 u_T^2 h_1}{v_1 v_2^{3/2} u_S} \,,\qquad
 H^{234} = - \frac{8 h_2}{v_1 v_2^{3/2} u_S \sqrt{g_3}} \,,
\nonumber \\
&& e^{-2\Phi} = u_S \,,\qquad\qquad T = u_T \,.
\end{eqnarray}
Here $g_3$ is a determinant of the metric on the unit 3-sphere $S^3$
(with coordinates $x^i$, $i=2,3,4$).

We follow the procedure from section \ref{sec:4chbh}. Lift of (\ref{d6d5}) to six 
dimensions gives \begin{eqnarray} \label{d5d6}
&& ds_6^2 \equiv G^{(6)}_{MN} dx^M dx^N
 = ds^2 + u_T^2 \left( dx^5 + 2\widetilde{e}_1 rdt \right)^2 \,,
\nonumber \\
&& K^{(6)}_{tr5} = \frac{\widetilde{e}_2}{2} \,, \qquad\qquad\qquad
 K^{(6)}_{234} = K_{234} = \frac{\widetilde{p}}{4} \sqrt{g_3} \,,
\nonumber \\
&& H^{(6)tr5} = \frac{4 h_1}{v_1 v_2^{3/2} u_S} \,, \qquad
 H^{(6)234} = - \frac{8 h_2}{v_1 v_2^{3/2} u_S \sqrt{g_3}} \,,
\nonumber \\
&& e^{-2\Phi^{(6)}} =  \frac{u_S}{8\pi\, u_T} \,.
\end{eqnarray}
Now $v_1$, $v_2$, $u_S$, $u_T$, $\widetilde{e}_1$, $\widetilde{e}_2$, $h_1$
and $h_2$ are unknown variables whose solution is to be found by extremizing
the entropy function for the fixed values of electric and magnetic charges 
$\widetilde{q}_{1,2}$ and $\widetilde{p}$. Entropy function is now given by
\begin{eqnarray} \label{entf5d}
\mathcal{E} &=& 2\pi \left( \sum_{i=1}^2 \widetilde{q}_i \, \widetilde{e}_i
 - \int_{S^3} \sqrt{-G} \, \widetilde{\mathcal{L}} \right)
= 2\pi \left( \sum_{i=1}^2 \widetilde{q}_i \, \widetilde{e}_i
 - \int_{S^3} \sqrt{-G^{(6)}} \, \widetilde{\mathcal{L}}^{(6)} \right)
\nonumber \\
&=& \mathcal{E}_0 + \mathcal{E}'_1 + \mathcal{E}''_1 \,,
\end{eqnarray}
where
\begin{eqnarray} \label{ef50}
\mathcal{E}_0 &=& 2\pi \left[ \widetilde{q}_1\widetilde{e}_1
 + \widetilde{q}_2 \widetilde{e}_2 - \frac{\pi}{16} v_1 v_2^{3/2} u_S
 \left( -\frac{2}{v_1} + \frac{6}{v_2}
 + \frac{2 u_T^2 \widetilde{e}_1^2}{v_1^2}
 + \frac{32\, h_2 (2\widetilde{e}_2 - h_2)}{v_1^2\, u_S^2}
 \right. \right.
\nonumber \\
&& \left. \left. \qquad
 - \frac{8 u_T^2 h (2\widetilde{p} - h)}{v_2^3\, u_S^2} \right)
 \right] \,,
\end{eqnarray}

\begin{equation} \label{ef51p}
\mathcal{E}_1' = - 2\pi^2 v_1 v_2^{3/2} u_S
 \Bigg[ \frac{512\, e_2 h_2^3}{v_1^4 u_S^4}
 + \frac{32\, u_T^4 \widetilde{p}\, h_1^3}{v_2^6 u_S^4}
 + \frac{8 u_T^4 \widetilde{p}\, h_1 \widetilde{e}_1^2}{v_1^2 v_2^3 u_S^2}
 - \frac{8 u_T^2 \widetilde{p}\, h_1}{v_1 v_2^3 u_S^2}
 - \frac{96\, \widetilde{e}_2 h_2}{v_1^2 v_2 u_S^2} \Bigg] \,,
\end{equation}

\begin{equation} \label{ef51pp}
\mathcal{E}_1'' = - 8\pi^2 \widetilde{p} \left(
 \frac{u_T^2}{v_1} \widetilde{e}_1
 - 2 \frac{u_T^4}{v_1^2} \widetilde{e}_1^3 \right) \,.
\end{equation}
Again, to obtain (\ref{ef51pp}) we had to deal with gravitational Chern-Simons term. However, 
this was already done in \cite{Cvitan:2007hu}, so we just copied the result.

The correspondence to the integer-valued charges $(n,w,m)$ appearing in string 
theory was discussed in \cite{Cvitan:2007hu}. The result was
\begin{equation} \label{3chnorm}
\widetilde{q}_1 = \frac{n}{2} \,,\qquad \widetilde{q}_2 = - 16\pi m
\,,\qquad \widetilde{p} = - \frac{w}{\pi} \,.
\end{equation}
Here $n$ and $w$ are momentum and winding number of string wound around $S^1$.
We postpone interpretation of $m$ for a moment.

Again, we were able to find analytic solutions to algebraic system for all values of
charges. For clarity of presentation, we restrict to $w,m>0$. Then $n>0$ ($n<0$) correspond 
to BPS (non-BPS) black holes. 

In the BPS case near-horizon solutions for 3-charge black holes
are given by
\begin{eqnarray} \label{3solb}
&& v_1 = 4(m+1) \,,\qquad v_2 = 4 v_1 \,,\qquad
 u_S = \frac{1}{8\pi} \sqrt{\frac{nw}{(m+1)(m+3)}} \,,\qquad
 u_T = \sqrt{\frac{n(m+1)}{w(m+3)}} \,,
\nonumber \\
&& \widetilde{e}_1 = \frac{1}{n} \sqrt{nw(m+3)} \,,\qquad
 \widetilde{e}_2 = h_2 = - \frac{1}{32\pi} \sqrt{\frac{nw}{m+3}} \,,\qquad
 h_1 = - \frac{w}{\pi} \,.
\end{eqnarray}
For the entropy we obtain
\begin{equation} \label{3entb}
S_{\rm bh}^{\rm BPS} = 2\pi \sqrt{nw(m+3)} \,.
\end{equation}

In the non-BPS case ($n<0$) we obtain
\begin{eqnarray} \label{3soln}
&& v_1 = 4(m+1) \,,\qquad v_2 = 4 v_1 \,,\qquad
 u_S = \frac{\sqrt{|n|w}}{8\pi(m+1)} \,,\qquad
 u_T = \sqrt{\frac{|n|}{w}} \,,
\nonumber \\
&& \widetilde{e}_1 = \frac{1}{n} \sqrt{|n|w(m+1)} \,,\qquad
 \widetilde{e}_2 = h_2 = - \frac{1}{32\pi} \sqrt{\frac{|n|w}{m+1}} \,,\qquad
 h_1 = - \frac{w}{\pi} \,.
\end{eqnarray}
For the entropy we obtain
\begin{equation} \label{3entn}
S_{\rm bh}^{\rm non-BPS} = 2\pi \sqrt{|n|w(m+1)} \,.
\end{equation}
This is exactly equal to the result conjectured in \cite{Cvitan:2007hu} (on the basis of
$\alpha'^3$-order perturbative results).

There is a subtle issue connected to interpretation of charges. Naively, we would expect that
charge $m$ should be equal to the number of NS5-branes, which we denote by $N$. To check
this, let us calculate components of 3-form strength $H^{(6)}$ with indices on $S^3$, 
evaluated on our solutions (\ref{3solb}) and (\ref{3soln}). The result is
\begin{equation} \label{H6S}
H^{(6)}_{234} = 32 (m+1) \sqrt{g_3} \,.
\end{equation}
From (\ref{H6S}) follows that $(m+1)$ is the magnetic charge (factor of 32 is from $2 \alpha'$).  As magnetic charges have topological origin, and so are not expected to receive perturbative corrections, 
we conclude that the number of NS5-branes should be given by
\begin{equation} \label{N5m}
N = m+1 \,.
\end{equation}
Using this in (\ref{3solb}) and (\ref{3soln}) we obtain our solutions expressed using "natural"
charges of the string theory, i.e., momentum $n$, winding $w$ and number of NS5-branes
$N$.

In the BPS case near-horizon solution (\ref{3solb}) becomes
\begin{eqnarray} \label{3solbp}
&& v_1 = 4 N \,,\qquad v_2 = 4 v_1 \,,\qquad
 u_S = \frac{1}{8\pi} \sqrt{\frac{nw}{N(N+2)}} \,,\qquad
 u_T = \sqrt{\frac{n N}{w(N+2)}} \,,
\nonumber \\
&& \widetilde{e}_1 = \frac{1}{n} \sqrt{nw(N+2)} \,,\qquad
 \widetilde{e}_2 = h_2 = - \frac{1}{32\pi} \sqrt{\frac{nw}{N+2}} \,,\qquad
 h_1 = - \frac{w}{\pi} \,,
\end{eqnarray}
while the entropy (\ref{3entb}) is
\begin{equation} \label{3entbp}
S_{\rm bh}^{\rm BPS} = 2\pi \sqrt{nw(N+2)} \,.
\end{equation}

In the non-BPS case near-horizon solution was given with (\ref{3solb}) which now becomes
\begin{eqnarray} \label{3solnp}
&& v_1 = 4 N \,,\qquad v_2 = 4 v_1 \,,\qquad
 u_S = \frac{\sqrt{|n|w}}{8\pi N} \,,\qquad
 u_T = \sqrt{\frac{|n|}{w}} \,,
\nonumber \\
&& \widetilde{e}_1 = \frac{1}{n} \sqrt{|n|wN} \,,\qquad
 \widetilde{e}_2 = h_2 = - \frac{1}{32\pi} \sqrt{\frac{|n|w}{N}} \,,\qquad
 h_1 = - \frac{w}{\pi} \,,
\end{eqnarray}
and the entropy (\ref{3entn}) is
\begin{equation} \label{3entnp}
S_{\rm bh}^{\rm non-BPS} = 2\pi \sqrt{|n|wN} \,.
\end{equation}
Though a detailed analyses of our near-horizon solutions (\ref{3solbp}) and (\ref{3solnp}) 
will be given in section \ref{sec:comm}, let us note here the following important properties:
\begin{itemize}
\item
Non-BPS solution (\ref{3solnp}) is $\alpha'$-\emph{un}corrected in our scheme. Now, it was 
shown that lowest-order BPS solution is an $\alpha'$-exact solution from the sigma model 
calculations \cite{Tseytlin:1996as} (corresponding result for 4-charge 4-dimensional black holes 
was given in \cite{Cvetic:1995bj}). As we use different scheme, our solutions cannot be directly
compared to sigma model ones. 
\item
The expressions for black hole entropies (\ref{3entbp}) and (\ref{3entnp}) are in agreement 
with those obtained from AdS/CFT correspondence, using the results for central charges 
calculated in \cite{Kutasov:1998zh} (see section \ref{ssec:cft} for more details).  
\end{itemize} 

Finally, it is easy to check that both BPS and non-BPS near-horizon solutions presented in this 
section satisfy 6-dimensional relation (\ref{modrie0}), which again means that inclusion of 
$\Delta\mathcal{L}^{(6)}_{\rm CS}$ in the action would not change our solutions and entropies 
(so they are also solutions of the action (\ref{6dtlsusy})).

\section{Comments on the solutions}
\label{sec:comm}

\subsection{AdS$_3$/CFT$_2$ correspondence}
\label{ssec:cft}

The solutions that we found and presented in sections \ref{sec:4chbh} and \ref{sec:3chbh}
are locally isomorphic to AdS$_3 \times S^3$ geometry. The quickest way to realize this is
to notice that all of them satisfy
\begin{eqnarray*} 
R^{(6)}_{MNPQ} &=& - \ell_A^{-2}
 \left( G^{(6)}_{MP} G^{(6)}_{NQ} - G^{(6)}_{MQ} G^{(6)}_{NP} \right)
 \quad \mbox{for} \quad M,N,P,Q \in \{0,1,y\} \\
&=& \ell_S^{-2} \left( G^{(6)}_{MP} G^{(6)}_{NQ} - G^{(6)}_{MQ} G^{(6)}_{NP} \right)
 \quad \mbox{for} \quad M,N,P,Q \in \{2,3,z\} \\
\end{eqnarray*}
where $\ell_A$ and $\ell_S$ play the role of radii of AdS$_3$ and $S^3$, 
respectively. For 4-dimensional 4-charge near-horizon solutions of section \ref{sec:4chbh}
we have $y=4$, $z=5$, and the radii are
\begin{equation} \label{4l2}
\ell_A^2 = \ell_S^2 = 16 (N'W'+2) \,,
\end{equation}
both for BPS and non-BPS solutions. For 5-dimensional 3-charge near-horizon solutions of 
section \ref{sec:3chbh} we have $y=5$, $z=4$, and the radii are
\begin{equation} \label{3l2}
\ell_A^2 = \ell_S^2 = 16 N \,,
\end{equation}
again the same for BPS and non-BPS solutions.

A presence of AdS$_3$ suggests that one can use powerful methods of AdS/CFT 
correspondence for obtaining the entropies. Using the dual CFT$_2$ formulation one obtains 
that the asymptotic expression for the black hole entropy should be given by Cardy formula 
\cite{cardy}
\begin{equation} \label{cardyf}
S_{\rm CFT} = 2\pi \sqrt{\frac{c_R \, n_R}{6}} + 2\pi \sqrt{\frac{c_L \, n_L}{6}} \,.
\end{equation}
For the extremal black holes that we analyzed, $n_R=0$ and $n_L=n$ in the BPS case, and
$n_L=0$ and $n_R=|n|$ in the non-BPS case. Generally, it is nontrivial to determine central
charges $c_R$ and $c_L$, but in the case of heterotic black holes analyzed in this paper the 
explicit sigma model calculations are possible and were done in \cite{Kutasov:1998zh}. In the 
case relevant for 4-charge black holes in $D=4$ the result was\footnote{The dual CFT$_2$ was 
proposed in \cite{Hohenegger:2008du}.}
\begin{equation} \label{c4ch}
c_R = 6 (N'W'+2) \,, \qquad c_L = 6 (N'W'+4) \,,
\end{equation}
while in the case relevant for 3-charge black holes in $D=5$
\begin{equation} \label{c3ch}
c_R = 6 N \,, \qquad c_L = 6 (N+2) \,.
\end{equation}
When (\ref{c4ch}) and (\ref{c3ch}) are plugged in Cardy formula (\ref{cardyf}) one obtains exactly 
the black hole entropies from sections \ref{sec:4chbh} and \ref{sec:3chbh}, i.e., (\ref{4entb}), 
(\ref{4entn}), (\ref{3entbp}) and (\ref{3entnp}).

Later it was shown \cite{Kraus:2005vz,Kraus:2006wn,David:2007ak} that when effective 
3-dimensional theory on AdS$_3$ has $(0,4)$ (or even smaller $(0,2)$ \cite{Kaura:2008us}) 
supersymmetry, central charges are generally determined purely by the coefficients of 
Chern-Simons terms. This method of calculating central charges has two virtues: (i) it is general,
depending only on symmetries, (ii) as Chern-Simons terms are connected to anomalies and 
correspondingly 1-loop saturated, their coefficients in many cases can be calculated exactly (at
least in $\alpha'$). In fact, in \cite{Kraus:2005vz} the power of this method was demonstrated
by calculating central charges (\ref{c4ch}) relevant for the entropy of 4-dimensional 4-charge 
black holes. The corresponding gravity calculation was done in \cite{Castro:2007sd}, by 
calculating $c_R + c_L$ from the $AdS_3 \times S^2$ solution of the effective 5-dimensional 
$R^2$ supergravity action constructed in \cite{Hanaki:2006pj} (obtained by supersymmetrization of
gravitational Chern-Simons term). As for the case relevant for 5-dimensional 3-charge black holes, 
i.e., (\ref{c3ch}), such calculations were not performed. For the gravity calculation, one needs
6-dimensional $R^2$ action (to find $AdS_3 \times S^3$ solutions) which is not fully known. 

Our method can be used to obtain the missing gravity confirmation for (\ref{c3ch}). The central 
charges $c_{R,L}$ can be calculated by using a generalization of the Sen's entropy function 
formalism to AdS$_3 \times S^l$ geometries (equivalent to "c-extremization" method
reviewed in \cite{Kraus:2006wn}).\footnote{In \cite{Garousi:2007zb,Shu:2008yd} different types of 
extension of entropy function formalism to general AdS$_k \times S^l$ geometries were discussed.} 
Starting from the same 10-dimensional supersymmetric action as before, in the case relevant 
for 4-dimensional 4-charge black holes (AdS$_3 \times S^2$ geometry) we obtain (\ref{c4ch}), 
while in the case relevant for 5-dimensional 3-charge black holes (AdS$_3 \times S^3$ geometry) we 
obtain (\ref{c3ch}). We shall present details of the calculation in separate publication 
\cite{PDPgef}. As in our calculation the only relevant higher-derivative part of the action is the 
one directly connected with Chern-Simons term, our results are (as expected) in agreement with 
Kraus-Larsen method \cite{Kraus:2005vz}.

\subsection{Comparison with previous analyses}
\label{ssec:prevan}

The four- and five-dimensional black holes considered in this paper (we denote our results 
by "susy10") have been analyzed previously in the literature by using two different types of 
four-derivative corrections to the lowest order (two-derivative) action: (a) $\mathcal{N}=2$ 
off-shell supersymmetric $R^2$ corrections directly constructed in $D=4$ and 5 dimensions 
(denoted here "susyD"), and (b) pure Gauss-Bonnet correction ("GB").  As the starting points 
of all of the mentioned calculations are mutually inequivalent actions which contain only parts of 
$\alpha'$-corrections of the full effective action of heterotic string (which has infinite 
expansion in $\alpha'$), it is interesting to compare the results.\footnote{For $D=4$ dimensional 
black holes "susy4" and "GB" results are reviewed in detail in \cite{Sen:2007qy}. For $D=5$ 
dimensional black holes "susy5" and "GB" results can be found in \cite{Cvitan:2007pk}.} 
Of course, to do this properly one would have to deal with freedom coming from regular 
field redefinitions\footnote{To properly take into account field redefinition freedom, one
needs to know the action fully up to particular order. The interesting discussion related to
this, in the context of AdS$_5 \times S^5$ solutions in type-IIB theory, can be found in 
\cite{Liu:2008kt}.} and gauge-fixings (indeed, one look at these solutions reveals that they 
are all mutually different). We shall restrict ourselves here to few, potentially interesting, 
remarks.

In $D=4$ dimensions, for BPS black holes all actions ("susy10", "susy4" and "GB") are 
leading to the same entropy formula, which agrees exactly in $\alpha'$ with statistical entropy 
(\ref{e4c4db}) obtained by counting of microstates in heterotic string theory. As for the 
near-horizon solutions, especially interesting is the similarity between our "susy10" (\ref{4solb}) 
and "GB" solutions (eqs. (3.1.56), (3.1.57) in \cite{Sen:2007qy}); they match for  $v_1$, $v_2$, 
$u_S$, $e_1$, and $H^{(6)MNP}$. In fact, they differ only for $u_1$ and $u_2$, which are in 
the "GB" solution given by
\begin{equation} \label{u12GB}
u_1 = \sqrt{\frac{n}{w}} \,, \qquad u_2 = \sqrt{\frac{W'}{N'}} \,.
\end{equation}
Now, we saw in section \ref{sec:4chbh} that $u_1$ and $u_2$ have naive interpretation as 
moduli (proper radii) $T_1$ and $T_2$ of the compactification circles.
With this interpretation, it is (\ref{u12GB}) which is consistent with T-dualities of the 
string theory ($T_1 \to 1/T_1$ and $n \leftrightarrow w$, $T_2 \to 1/T_2$ and 
$N' \leftrightarrow W'$), unlike $u_1$ and $u_2$ from our "susy10" solution (\ref{4solb}). 
Our analysis strongly suggest that, not only that "GB" near-horizon solution should be taken 
seriously, and is probably correct, but also that it is more directly connected to stringy 
geometry.\footnote{The fact that $u_1$ and $u_2$ of "susy10" solution do not respect naively 
implemented T-dualities does not mean that solution is wrong, but that probably one needs field redefinitions to connect them to moduli $T_1$ and $T_2$ of the stringy geometry. It is not 
unusual that inclusion of higher-derivative corrections in the action induce corrections to 
physical interpretations of fields (for explicit example see \cite{Sen:2004dp}).} We emphasize 
this because the fact that simple Gauss-Bonnet correction leads to $\alpha'$-exact agreement 
with statistical entropy formula is still not understood well, and is appearing almost as 
a miracle.

As for "susy4" near-horizon BPS solution, it was shown in \cite{Sen:2005iz} that it almost 
matches with "GB", differing only in $u_1$ and $u_2$ \cite{Sen:2007qy}. This mismatch is again
probably due to the different field redefinition schemes. Altogether, we tend to believe that
"susy10", "susy4" and "GB" near-horizon solutions for BPS 4-dimensional 4-charge black holes 
are all equivalent. 

For non-BPS black holes in $D=4$ only our "susy10" solution is giving correct statistical entropy
(\ref{e4c4dn}). Both "susy4" and "GB" solutions are giving wrong results already at $\alpha'^1$ 
order, which signals that corresponding actions are incomplete already at four-derivative level 
(but for some unknown reason are giving exact results for BPS black holes) \cite{Sen:2007qy}.   

In $D=5$ dimensions, for large ($N\ne0$) BPS black holes only "susy10" and "susy5" are 
leading to the entropy formula (\ref{e3c5db}), which is in agreement with prediction 
based on AdS/CFT conjecture \cite{Kutasov:1998zh} (direct stringy statistical calculation is still 
not known). "GB" entropy differs starting from $\alpha'^2$-order \cite{Cvitan:2007pk}. 
Again, it is interesting to compare our near-horizon solution "susy10" with "susy5", given in Eq. 
(5.28-30) of \cite{Cvitan:2007en} (with $\zeta=1$). After passing to the string frame, "susy5"
solution becomes\footnote{In obtaining expression for $u_T$ in (\ref{susy5b}) we used 
$T = (M^1)^{-1/2} (M^2)^{-1}$ (see \cite{Cvitan:2007pk} for notations), relation valid in the 
two-derivative approximation. But, in this approximation, from the on-shell condition
for prepotential $\mathcal{N} \equiv M^1 M^2 M^3 = 1$ (real special geometry) follows
that one could use also $T = (M^1)^{1/2} M^3 $ or $T = (M^3/M^2)^{1/2}$. That, however, give 
different expressions for $u_T$, which is a consequence of  the fact that higher-derivative 
corrections make $\mathcal{N} \ne 1$. For the choice $T = (M^3/M^2)^{1/2}$ one gets
$u_T = \sqrt{n/w}$, which means that this could be a correct identification with heterotic 
string compactification modulus (radius of $S^1$) \cite{Cvitan:2007pk}. Similarly, we 
obtained $u_S$ in (\ref{susy5b}) by using $S = (M^1)^{3/2}$, instead of $S = (M^2 M^3)^{-3/2}$, 
which is equivalent in the two-derivative approximation, but receives different higher 
$\alpha'$ corrections.}
\begin{eqnarray} \label{susy5b}
&& v_1 = \frac{\alpha'}{4}(m+1) \,,\quad v_2 = 4 v_1 \,,\quad
 u_S = \frac{\pi\alpha'^{-3/2}}{4 G_5} \sqrt{\frac{nw}{(m+1)(m+3)}} \,,\quad
 u_T = \sqrt{\frac{n(m+1)}{w(m+3)}} \,,
\nonumber \\
&& e_1 = \frac{1}{2 \alpha'} \sqrt{\frac{nw}{(m+3)}} \,,\qquad
 e_2 = \frac{\sqrt{\alpha'}}{2n} \sqrt{nw(m+3)} \,,\qquad
 e_3 = \frac{\sqrt{\alpha'}}{2w} \sqrt{nw(m+3)} \,.
\end{eqnarray}
In \cite{Cvitan:2007en} different conventions were used ($\alpha'=1$, $G_5=\pi/4$). Using
conventions from the present paper, which include $\alpha'=16$ and $G_5=2$, but also 
transformations on gauge fields which include passing from 3-form $K$ to $H$ (i.e., "removing" 
tildes), renaming of indices, additional factors of 2 coming from different normalization in the 
corresponding actions, and finally $N=m+1$, it is easy to show that "susy5" solution 
(\ref{susy5b}) becomes \emph{exactly} our "susy10" solution (\ref{3solbp}). This matching is 
not that surprising, considering that the starting actions were both obtained by
supersymmetrization of gravitational Chern-Simons term, in one case in $D=10$ (on-shell
supersymmetry) and in the second case directly in $D=5$ (off-shell supersymmetry).

For non-BPS black holes in $D=5$, "susy5" and "GB" solutions are giving results
for the entropy disagreeing with "susy10" entropy (\ref{3entnp}) already at $\alpha'^1$-order. 
As (\ref{3entnp}) can be obtained from (\ref{c3ch}) by using AdS/CFT arguments 
\cite{Cvitan:2007hu}, this again signals that "susy5" and "GB" actions are incomplete (as 
heterotic string effective actions) already at four-derivative level.

\subsection{Additional terms in the action}
\label{ssec:torsion}

We have obtained results for black hole entropies which match statistical entropies of string 
theory exactly in $\alpha'$ by using part of the effective action obtained by supersymmetrization 
of the gravitational Chern-Simons term. As we already discussed in section \ref{sec:eahet}, this 
is not a complete action, and there are other terms (denoted by 
$\mathcal{L}^{(10)}_{\mathrm{other}}$ in (\ref{6dtl})) starting from $\alpha'^3$-order. 
Natural question is: why these other terms do not contribute? One explanation comes from 
AdS$_3$/CFT$_2$ (plus $(0,4)$ supersymmetry) anomaly inflow arguments \cite{Kraus:2005vz} mentioned in section \ref{ssec:cft}: in 3-dimensional language the only relevant terms are 
Chern-Simons terms. So, it is natural to expect also from 10-dimensional perspective that the 
only terms which are important are Chern-Simons terms and terms connected to them by 
supersymmetry. However, we find it interesting to address the above question directly, as it can 
give us some new information on the structure of the 10-dimensional effective action.

From the fact that our calculation was successful for different types of black holes, it is
natural to assume that cancellations appear because of some general property of solutions. In
fact, even before inclusion of $\mathcal{L}^{(10)}_{\mathrm{other}}$, we have seen in section 
\ref{ssec:10Daction} that we were able to handle infinite number of terms in 
$\Delta \mathcal{L}^{(10)}_{\mathrm{CS}}$ because of the following properties:
\begin{enumerate}
\item
Every term in $\Delta \mathcal{L}^{(10)}_{\mathrm{CS}}$ has at least two
powers of Riemann tensors $\overline{R}^{(10)}_{MNPQ}$, calculated using connection 
with torsion (\ref{modcon}). (In fact, every monomial at $\alpha'^n$ order is obtained by 
contraction of $(n+1)$ Riemanns $\overline{R}^{(10)}_{MNPQ}$
\cite{Bergshoeff:1989de}.)
\item
Neglecting $\Delta \mathcal{L}^{(10)}_{\mathrm{CS}}$, obtained near-horizon 
solution satisfies $\overline{R}^{(10)}_{MNPQ} = 0$.
\end{enumerate}
From these properties it trivially follows that $\Delta \mathcal{L}^{(10)}_{\mathrm{CS}}$ is 
giving vanishing contribution to near-horizon equations of motion and black hole entropy.

Now, if the property 1 would hold also for $\mathcal{L}^{(10)}_{\mathrm{other}}$ (weak form 
of the conjecture) the same reasoning would immediately prove that this term is irrelevant in 
calculations of near-horizon solutions and the entropies. In fact, property 1 was conjectured long 
time ago in \cite{Metsaev:1987zx}. After explicit 4-point level calculations 
\cite{Gross:1986mw,Policastro:2006vt} confirmed this by showing that at this level 
$\mathcal{L}^{(10)}_{\mathrm{other}}$ can be constructed just from monomials which are pure 
contracted products of $\overline{R}_{MNPQ}$ (a stronger version of the conjecture), the 
conjecture was taken more seriously and used (in the stronger form) in literature, see, e.g., 
\cite{Kehagias:1997cq}. 

However, results of more recent calculations of 1-loop 5-point amplitudes in type-IIB string 
theory\footnote{The 1-loop part of the NS-NS sector of type IIB effective action has the same 
form as the tree-level part, which is equal for all superstring theories.} in 
\cite{Frolov:2001xr,Peeters:2001ub} appeared to violate the stronger version of conjecture. As, 
in addition, no one has found any convincing argument why the conjecture (in weaker or 
stronger form) should be correct, a widespread opinion among the experts was that it is indeed 
wrong.\footnote{I am grateful to K.\ Peeters for discussions on this point.}

And then, the most recent detailed calculation \cite{Richards:2008sa} of the 1-loop 
5-point type-IIB amplitudes gave results which are \emph{in agreement with the conjecture 
(stronger form)}. Now, how this new twist can be compatible with the results from 
\cite{Frolov:2001xr,Peeters:2001ub}? Calculations in \cite{Peeters:2001ub}, as already
noted by the authors, were incomplete, because subtractions due to quartic terms in the
action were not done. A disagreement with \cite{Frolov:2001xr} is more mysterious.

If the conjecture, as supported by \cite{Richards:2008sa}, is correct (at least in weaker form,
obtained by replacing $\Delta \mathcal{L}^{(10)}_{\mathrm{CS}}$ with
$\mathcal{L}^{(10)}_{\mathrm{other}}$ in property 1 above), then we are guaranteed that our 
near-horizon solutions from sections \ref{sec:4chbh} and \ref{sec:3chbh} are indeed 
\emph{$\alpha'$-exact solutions of the full 10-dimensional heterotic effective action}.

Starting from the opposite side, the fact that our near-horizon solutions are giving the black hole 
entropies which are $\alpha'$-exactly equal to microscopic statistical entropies, implies
that our calculations support the above conjecture (at least in the weak form) on the form of
$\mathcal{L}^{(10)}_{\mathrm{other}}$. Though, because of the simplicity of our solutions (e.g., 
all covariant derivatives vanish), results of the present paper are insufficient to prove the 
conjecture, they can be used to extract interesting relations between terms of the form 
$R^k H^{2l}$ in the effective action.

At the end, let us mention one, almost trivial, consequence of the conjecture. Let us
consider the same type of black holes, but now in type II string theories (compactified on the 
same manifolds as before). As for these theories the only relevant $\alpha'$-corrections are 
given by $\mathcal{L}^{(10)}_{\mathrm{other}}$, which is the same as in heterotic action, 
we obtain that the near-horizon solutions and entropies stay \emph{uncorrected}.
This means that the entropy formula for 4-charge black holes in $D=4$ (compactification
on $S^1 \times S^1 \times T^4$) is
\begin{equation} \label{4entII}
S_{\rm bh} = 2\pi \sqrt{|nwN'W'|} \,,
\end{equation}
while for 3-charge black holes in $D=5$ (compactification on $S^1 \times T^4$) is
\begin{equation} \label{3entII}
S_{\rm bh} = 2\pi \sqrt{|nwN|} \,.
\end{equation}
For the large BPS black holes (when all charges are nonvanishing), (\ref{4entII}) and 
(\ref{3entII}) follow from OSV conjecture \cite{Ooguri:2004zv}. It would be interesting to find 
out could the above argument be used for more general black holes in type II string theories, 
like those analyzed in \cite{Cai:2006xm}.

\subsection{Small black holes}
\label{ssec:smallBH}

When one takes magnetic charges, which are $N'$ and $W'$ for 4-charge states analyzed
in section \ref{sec:4chbh}, and $N$ for 3-charge states analyzed in section \ref{sec:3chbh}, 
to \emph{vanish}, one obtains 2-charge states describing a fundamental heterotic string on 
$M_D \times S^1 \times T^{9-D}$ (where $D=4$ or 5, respectively) having a momentum $n$ and 
winding number $w$ on $S^1$. These, so called Dabholkar-Harvey states, can be defined for any 
$D \le 9$. For such states, which are pure perturbative 
string states, it is easy to calculate statistical entropy asymptotically for $|nw|\gg1$. 

For BPS states which satisfy $n,w>0$ statistical entropy is
\begin{equation} \label{2entb}
S_{\rm stat}^{\rm BPS} = 4\pi \sqrt{nw} \,,
\end{equation}
while for non-BPS states satisfying $n<0$, $w>0$ statistical entropy is
\begin{equation} \label{2entn}
S_{\rm stat}^{\rm non-BPS} = 2\sqrt{2} \pi \sqrt{|nw|} \,.
\end{equation}

On the gravity side, these states should correspond to \emph{small} black hole solutions 
(see \cite{Cornalba:2006hc} and references therein) which at the lowest order (Einstein gravity) 
have singular horizon with vanishing area (and entropy). Inclusion of $\alpha'$-corrections is
expected to make horizon regular, but with the radius of the order of string length. This means that curvature scalars are of the order $1/\alpha'$, which suggests that for such solutions low energy 
(small curvature) effective action should not be useful. This was explicitly shown in 
\cite{Sen:2005kj}, where higher-curvature corrections were modeled by simple Gauss-Bonnet term 
("GB"-action, in the language of sec. \ref{ssec:prevan}). Surprisingly, this simple action gives the 
entropy which agrees with statistical result in the BPS case (\ref{2entb}) in $D=4$ and 5 (but fails to reproduce (\ref{2entn}) in the non-BPS case). We mention that the BPS entropy formula 
(\ref{2entb}) can be reproduced in all dimensions if one takes unique action whose 
$\alpha'$-correction is purely given by extended Gauss-Bonnet densities \cite{Prester:2005qs}, but 
it is still unclear why such action should be relevant. 

Let us analyze first small black hole limit in $D=4$, which is obtained by taking $N',W' \to 0$ (by 
keeping $N'/W'$ fixed and finite) in 4-charge near-horizon solutions (\ref{4solb}) and (\ref{4soln}). 
For the AdS$_2$ and $S^2$ radii we obtain $r_A = r_S = \sqrt{8} = \sqrt{\alpha'/2}$. The limit is 
regular for all variables, except for $u_2$ (term inside round bracket obviously diverges). At first, 
this appears as a serious problem, because $u_2$ should correspond to proper radius  $T_2$ 
(measured in $\alpha'$-units, see eq. (\ref{volum})) of one of the compactification circles. But, from
T-duality we expect to have $T_2 = \sqrt{W'/N'}$, which is differing from expressions for $u_2$ in
(\ref{4solb}) and (\ref{4soln}) exactly by the problematic term in round brackets. Now it is obvious 
what is happening here. Higher-derivative corrections have changed the physical meaning of 
$u_2$, and to get back to the standard interpretation one needs a field redefinition. It appears 
here that the requested field redefinition, which should remove round bracket in expression for 
$u_2$, is  \emph{singular} in the small black hole limit. Though the field redefinition analysis 
appears quite tricky\footnote{One can easily find simple field redefinition which does the job for 
$u_2$, for example
\begin{equation}
T_2 = u_2 \sqrt{1 - 16\,u_2^2 \left(F^{(2)}_{\mu\nu}\right)^{\!2}} \;,
\end{equation}
but this by itself will not remove all the tricky terms in the entropy function (which behave 
non-trivially in the small black hole limit).}, and goes beyond the present paper, we expect that 
obtained small black hole solutions are meaningful. This is supported by the results for the entropy: 
taking $N',W' = 0$ in (\ref{4entb}) and (\ref{4entn}) is giving exact agreement with statistical 
entropies (\ref{2entb}) and (\ref{2entn}). Let us mention that "susy4"-action (as "GB"-action) is 
reproducing the statistical entropy only in BPS case (\ref{2entb}) 
\cite{Sen:2004dp,Dabholkar:2004dq,Hubeny:2004ji}.

Let us now see what is happening for 3-charge small black holes in $D=5$. Plugging $N=0$ in 
near-horizon solutions (\ref{3solbp}) and (\ref{3solnp}) one gets \emph{completely singular}
solutions, where AdS$_2$ and $S^3$ radii, modulus $u_T$ (BPS case), and effective string 
coupling ($1/\sqrt{u_S}$) all vanish. It appears that $\alpha'$ corrections considered here are not regularizing the horizon. In this case taking naive limit $N\to0$ in black hole entropy formulae 
(\ref{3entbp}) and (\ref{3entnp}) is meaningless, so it is not strange that the small black hole limit 
does not give statistical entropies (\ref{2entb}) and (\ref{2entn}). All our efforts, analytical and 
numerical, for finding regular AdS$_2 \times S^3$ solutions from Lagrangians (\ref{6dtlred}) and 
(\ref{6dtlsusy}) failed. 

Let us mention that the same happens for corresponding small black holes in type-II theories, 
for which statistical entropy, for all values of charges, is given by (\ref{2entn}). Near-horizon 
solutions and large black hole entropies are $\alpha'$-uncorrected, both in $D=4$ and $D=5$,
and so obviously have completely singular small black hole limits. 

There are three possible explanations:
\begin{enumerate}
\item
Horizon geometry drastically changes. 
\item
Low energy effective action is useless for such small black holes (as naively expected). It may be
that this is just the problem of the scheme used, and some singular field redefinition could put
the action in the form which has regular small black hole solutions. Our analysis suggests 
that these (singular) field redefinitions should be much more complicated than those
needed in heterotic $D=4$ cases.
\item
For such small black holes new physically acceptable near-horizon solutions (of non-linear 
equations) with AdS$_2 \times S^{D-2}$ appear. If this is the case, our analysis shows that such 
solutions should not satisfy $\overline{R}_{MNPQ} = 0$. But, without this condition, we 
are unable to perform calculations because the effective actions are unknown (and also have 
infinite number of terms). 
\end{enumerate}
It would be interesting to understand the connection between our results, in particular the
difference between four and five-dimensional small black holes, and the analyses of fundamental 
string based on supersymmetry and holography 
\cite{Dabholkar:2007gp,Lapan:2007jx,Kraus:2007vu,Alishahiha:2007ap,Alishahiha:2008kc,Duff:2008pa,Hung:2007su,Hohenegger:2008du}.

\acknowledgments

We would like to thank L.\ Bonora, G.\ Lopes Cardoso, K.\ Peeters and B.\ Sahoo for valuable 
discussions. This work was supported by the Croatian Ministry of Science, Education and 
Sport under the contract No.\ 119-0982930-1016. P.D.P. was also supported by 
Alexander von Humboldt Foundation.

\end{document}